\documentclass[%
 amsmath,amssymb,
 aps, physrev,
pra,
]{revtex4-2}

\usepackage{graphicx}
\usepackage{dcolumn}
\usepackage{bm}

\begin{document}

\preprint{APS/123-QED}

\title{\textbf{Enhancing high-order harmonic generation in two-color laser fields: A comparison of single- and two-color schemes} 
}%

\author{Jan Tro\ss}
\affiliation{%
 J.R. Macdonald Laboratory, Department of Physics, Kansas State University, Manhattan, KS 66506, USA
}\author{Travis Severt}%
\affiliation{%
 J.R. Macdonald Laboratory, Department of Physics, Kansas State University, Manhattan, KS 66506, USA
}\author{Georgios Kolliopoulos}%
\affiliation{%
 J.R. Macdonald Laboratory, Department of Physics, Kansas State University, Manhattan, KS 66506, USA
}\author{Pratap Timilsina}%
\affiliation{%
 J.R. Macdonald Laboratory, Department of Physics, Kansas State University, Manhattan, KS 66506, USA
}\author{Itzik Ben-Itzhak}%
\affiliation{%
 J.R. Macdonald Laboratory, Department of Physics, Kansas State University, Manhattan, KS 66506, USA
}\author{Carlos A. Trallero-Herrero}%
\affiliation{%
 J.R. Macdonald Laboratory, Department of Physics, Kansas State University, Manhattan, KS 66506, USA
}%
\affiliation{%
 Department of Physics, University of Connecticut, Storrs, CT 06269-3046, USA
}
\date{\today}

\begin{abstract}
We compare the enhancement of high-order harmonic generation in argon using 800-400 and 800-266~nm laser fields with their optimized single-color counterparts. We observe that the two-color fields generally outperform the single-color 800~nm field by factors of 2 to 3800, depending on the photon energy and generation scheme. From this comparison, we determine which scheme is optimal for each photon energy. We also observe that the divergence of HHG produced by the 800-266~nm laser fields is smaller than that of the other single- and two-color driving fields, perhaps suggesting that the 800-266~nm fields optimize the recombination probability of the so-called ``short" electron trajectories.
\end{abstract}

\maketitle

\section{Introduction}

High-order harmonic generation (HHG) is an important technique, allowing researchers to produce vacuum-ultraviolet to soft X-ray laser pulses using table-top setups. Although large synchrotron and free-electron laser (FEL) facilities provide pulse energies higher than conventional HHG based laser sources, the laser pulses generated via HHG are capable of producing attosecond pulses with excellent spatial and temporal coherence properties~\cite{Salieres1999,Bellini1998,Bartels2002,Benko2014,Terschluesen2014}, which only state-of-the-art FEL facilities are beginning to approach~\cite{Huang2017,Coffee2019,Duris2019}. To make HHG-based laser sources promising tools for a variety of fields of study, research must be devoted to increasing the overall photon flux to decrease data acquisition times and improve signal-to-noise ratios.

Since HHG typically has low conversion efficiencies of the order of $10^{-5}$ or lower~\cite{Chang2011,Lin2018}, researchers have studied a variety of methods to increase photon flux. Some researchers use a brute force approach, where they develop high-power driving lasers to increase the photon flux and/or harmonic pulse energy on target~\cite{Takahashi2002,Haedrich2016,Heyl2017,Makos2020}.  Alternatively, others try to increase the efficiency of HHG by improving phase matching conditions, for example, by using loose focusing geometries in conjunction with gas cells~\cite{Tamaki1999,Sutherland2004,Rudawski2013,Sun2017} or gas-filled waveguide~\cite{Rundquist1998,Pfeifer2005,Paul2006,Popmintchev2009}. Further approaches explore changing the frequency of the driving field to increase the overall efficiency of HHG, for example, using shorter wavelength~\cite{Adachi2012,Popmintchev2015,Wang2015a,Marceau2017,Adachi2018} or multicolor driving fields~\cite{Watanabe1994,Andiel1999,Cormier2000,Kim2005,Liu2006,Kim2008,Lambert2009,Brugnera2010,Siegel2010,Wei2013,Brizuela2013,Wei2014,Jin2014,Jin2014a,Haessler2014,Jin2015,Kroh2018,Sayrac2019,raabXUVYieldOptimization2025}. More recently, there has also been interest in multicolor fields for optimizing HHG in fibers \cite{cirmiOpticalWaveformSynthesis2023, liGenerationIntenseLowDivergence2022}. 

In this paper, we focus on increasing the overall efficiency of the generated harmonics using two-color $\omega-2\omega$ and $\omega-3\omega$ driving laser fields, to provide a full context of comparison to our previous work \cite{severtEnhancingHighorderHarmonic2021}. Increasing the efficiency or flux of high-harmonic generation using two-color $\omega-2\omega$ laser fields has already been studied extensively in the past~\cite{Andiel1999,Cormier2000,Kim2005,Liu2006,Kim2008,Lambert2009,Brugnera2010,Siegel2010,Sayrac2019}, while using $\omega-3\omega$ has only recently begun to be explored in detail~\cite{Jin2014,Jin2014a,Jin2015,Kroh2018}. For example, Kim \emph{et al.}~\cite{Kim2005} showed that high-order harmonics produced by orthogonally polarized 800-400~nm laser fields were more than two orders of magnitude stronger than from the fundamental driving field alone, reporting a conversion efficiency as high as $5\times 10^{-5}$ for the 38th harmonic at 21.6~nm. Later, using the same approach, Kim \emph{et al.}~\cite{Kim2008} increased the efficiency of the 38th harmonic by extending the length of their gas jet to 6 mm, reaching pulse energies of 0.6 $\mu$J with a conversion efficiency as high as $2\times 10^{-4}$. More recently, Jin \emph{et al.}~\cite{Jin2014,Jin2014a} theoretically showed that the efficiency of high-order harmonic generation can be improved by two or more orders of magnitude using an $\omega-3\omega$ driving field. Their argument is that in the two-color $\omega-3\omega$ fields, the relative phase between the fields is tuned to optimize the recombination probability of the ``short trajectory" electrons.  Following their theoretical predictions, Kroh \emph{et. al.}~\cite{Kroh2018} systematically studied the enhancement of HHG driven by 2100($\omega$) and 700~nm ($3\omega$) driving fields, where they observed efficiency enhancements of 8.2 times in photon flux integrated from 20-70~eV and up to 2.2 times from 85-205~eV over the single-color 2100~nm laser field.

In order to take advantage of the wavelength scaling of HHG efficiency using ultraviolet driving fields~\cite{Adachi2012,Popmintchev2015,Wang2015a,Marceau2017,Adachi2018,Comby2019}, we study HHG driven by two-color 800-400~nm ($\omega-2\omega$) and 800-266~nm ($\omega-3\omega$) laser fields to further improve HHG efficiency from 15 to 40~eV when compared to the single-color 800~nm case. Additionally, we compare the relative HHG efficiency of the two-color driving fields with the same photon energies produced by the single-color 400 and 266~nm fields to determine whether it is worthwhile to use the more demanding two-color setup to generate certain photon energies.

It is worth noting that we take a pragmatic experimentalist approach for comparing the relative efficiencies between the optimized single- and two-color driving fields. Since laser systems are typically limited in their maximum output power, we chose to fix the maximum power input into our experimental setup instead of, for example, the total pulse energies or peak intensities in the target medium. Therefore, due to the efficiency of generating the second and third harmonics of the fundamental 800~nm driving field, the two-color and single-color 400 and 266~nm driving fields do not provide as much power on target as the fundamental driving field. Furthermore, we change phase-matching conditions, though admittedly in a limited range, for the different driving fields so we can compare the efficiency for the optimized conditions of each of our measurements, even though the optimized two-color and single-color 400 and 266~nm fields have lower input power, they generally outperform the 800~nm driving pulse.

\section{Experimental setup and methods}
The goal of this experiment is to compare harmonics produced from two-color 800-400 and 800-266~nm laser fields to their optimized single-color counterparts. In this section, we briefly describe how we generate and measure the flux of the harmonics in the optimized single-color and two-color laser fields.

We begin with a Ti:Sapphire laser that provides a pulse duration of 27 fs, full width-half maximum intensity, and a maximum pulse energy of 20 mJ at a repetition rate of 1 kHz~\cite{Langdon2015}. Since we are using an in vacuum $f=37.5$-cm spherical mirror to focus the laser onto the gas jet, we attenuate the pulse energy using beam splitters before our experimental apparatus to limit the peak intensity in our target. To have variable attenuation of our input power, we reflect the 800~nm laser pulses through a set of Germanium plates at Brewster's angle for 800 nm, providing a maximum pulse energy of 2 mJ before our experimental setup.

\begin{figure}[h]
 \centering
        \includegraphics[width=0.75\textwidth]{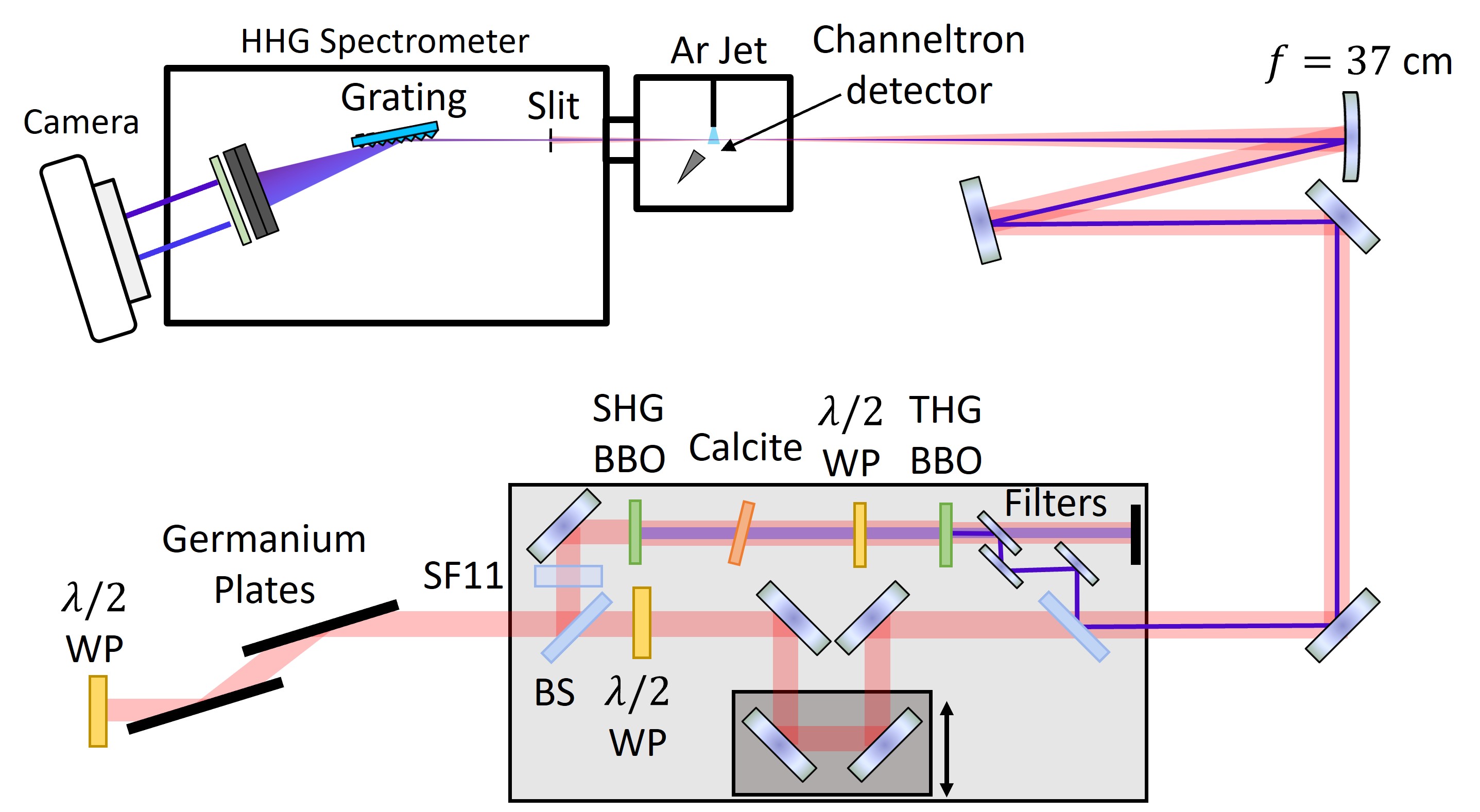}
    \caption{Schematic of the experimental setup (see text for description). The setup is the same as the one used in our previous work but here we make use of both SHG and THG for the two color synthesis. In addition to the ultrafast laser pulse there is an additional green laser on top of the interferometer providing an independent measurement of the interferometer without feedback. The root-mean-square error of the time-stability is of 50 as over 170 seconds. Adapted from \cite{severtEnhancingHighorderHarmonic2021}.}
    \label{fig:expsetup}
\end{figure}

When generating harmonics using the optimized 800~nm laser field, we directly focus the laser pulse on an argon gas jet, introduced through a glass capillary with an inner diameter of 250~$\mu$m. The backing pressure of the argon gas jet was kept constant at approximately 1100 Torr throughout the measurements, raising the pressure in the vacuum chamber from $1\times 10^{-8}$ to $1\times 10^{-5}$ Torr. To change phase matching conditions, we scan the gas jet $\pm 2.5$~cm throughout the focal profile while monitoring the relative ionization rate in the medium using a channeltron detector.

The produced harmonics then propagate through a differentially pumped region to a spectrometer, with an entrance slit of 450~$\mu$m. The harmonics then diffract off a flat-field VUV grating and are dispersed onto an MCP-phosphor detector. Then a camera collects the resulting fluorescence off the phosphor screen. To determine the yield of each harmonic, we integrate the fluorescence captured by the camera for 2 seconds. We also corrected the camera images for the polarization-sensitive efficiency of the grating, the wavelength-dependent detection efficiency of the microchannel plates~\cite{Martin1982}, and the nonuniform angular acceptance due to the spectrometer entrance slit.

For generating the optimized harmonics driven by a 400~nm laser field, we frequency double the 800~nm light after the germanium plates using a 15-mm diameter, 250$\mu$m thick $\beta$-BBO crystal (SHG-BBO), with the cut angle chosen to optimize second harmonic generation of 800~nm light for type-I phase matching. Note that the 400~nm light is orthogonally polarized to the 800~nm field. After filtering out the 400~nm light using four reflective dichroic beam splitters, we have a conversion efficiency of 40\%, producing a maximum of 880~$\mu$J of second harmonic. Since we do not compensate for the dispersion due to the 1-mm thick UV fused silica (UVFS) entrance window into the vacuum chamber, the 400~nm light is positively chirped with a pulse duration of 44 fs in the interaction region. The pulse duration was measured outside the vacuum chamber using a home-built self-diffraction FROG~\cite{Trebino1997} after passing through an equivalent 1\~mm thick UVFS blank.

To produce the 266~nm laser field, we send the orthogonally polarized 800 and 400~nm fields directly after the SHG-BBO through a 0.25~mm thick calcite, compensating for the delay between the 800 and 400~nm pulses. Then, the two-color beam passes through a zero-order $\lambda/2$ waveplate, which rotates the 800~nm light to the same polarization direction as the 400~nm light. The beam then propagates through a type-I BBO crystal (THG-BBO) that is 0.1~mm thick with an optimized cut angle for sum-frequency generation between the 800 and 400~nm pulses. Note that the resulting 266~nm field has the same polarization as the 800~nm field before the initial SHG-BBO. We finally separate the third harmonic from the fundamental and second harmonic using reflective dichroic beam splitters. After filtering, the third harmonic conversion efficiency is ~12\% of the fundamental, leading to a maximum pulse energy of 310~$\mu$J. The pulse duration of the 266~nm light was measured to be 59 fs, positively chirped.

To generate the $\omega-2\omega$ and $\omega-3\omega$ laser fields, we introduce a two-color interferometer immediately following the germanium plates, as shown in Fig.~\ref{fig:expsetup}(a). To avoid depletion effects in the fundamental driving field, we generate the second and third harmonics in one arm of the interferometer instead of before it, and therefore significantly improve the spatial quality of the beams produced by HHG. Explicitly, the 800~nm laser beam enters the interferometer and is split using a beam splitter. To control the relative intensities between the fundamental and either second or third harmonics, we use beam splitters with different reflection/transmission ratios.

The 800~nm beam transmitted through the beam splitter propagates through the delay arm of the interferometer, noted as ``Arm A" in Fig.~\ref{fig:expsetup}. The beam is transmitted through a $\lambda / 2$ waveplate, which only rotates the polarization of light by 90$^\circ$ when studying the $\omega-2\omega$ driving field. Since we do not need to rotate the polarization for the $\omega-3\omega$ measurements, we tune the waveplate to maintain the initial polarization direction. Note that the wave plate remained in the interferometer to ensure that the group-velocity-dispersion (GVD) of the 800~nm pulse is the same for the $\omega-2\omega$ and $\omega-3\omega$ measurements. Then, the 800~nm light reflects off a retro reflector on a piezo-controlled linear stage, which controls the relative time-delay/phase between the fields. Finally, the beam transmits through a dichroic beam splitter, reflects off several steering mirrors, and enters the vacuum chamber where the harmonics are produced and measured. Note that the dispersion of the 800~nm beam is optimized to produce the largest cutoff photon energy, the brightest harmonics, and the highest ionization rate measured by the channeltron detector simultaneously.

Meanwhile, in the other arm of the interferometer, labeled as ``Arm B" in Fig.~\ref{fig:expsetup}, the 800~nm beam reflects off the first beam splitter and transmits through a 2~mm thick SF11 glass and the SHG-BBO. We use SF11 glass to compensate for the negative GVD such that the second harmonic is produced with the optimal conversion efficiency. To generate the third harmonic, we placed the additional optics described above after the SHG-BBO, as shown in Fig.~\ref{fig:expsetup}. The generated second or third harmonic then reflects from four dichroic beam splitters to remove the fundamental 800~nm. After the final beam splitter, the second or third harmonic is spatially and temporally recombined with the 800~nm beam that propagated through ``Arm A." Finally, the two-color beam propagates into the vacuum chamber, where light is focused onto the argon gas jet.


For our two-color measurements to be successful, we need to have stability between the two arms that is ``much" smaller than the period of the shortest wavelength driving field, which is 0.9~fs for the 266~nm field. To achieve this stability, we vibrationally isolated the interferometer from the optics table. We measure the stability by propagating a frequency-doubled continuous wave ND:YAG laser through the interferometer and image the spatial interference at the exit. In \cite{severtEnhancingHighorderHarmonic2021}, we give details on the stability measurements of the interferometer showing a root mean square error (RMSE) of $\pm 50$~as over 170~seconds, which is indeed significantly smaller than the period of the 266~nm laser pulse.

\section{Optimized single-color driving fields}
In this section, we compare the harmonics produced by the optimized single-color driving fields. Due to the wavelength scaling of HHG efficiency for driving fields of 800~nm and shorter~\cite{Marceau2017}, we expect the HHG photon flux to improve significantly with shorter wavelength driving fields. Driving high-order harmonics using single-color laser fields may have some advantages over the two-color counterparts. Firstly, a single-color optical setup is significantly easier to design and build than a phase-stabilized two-color setup, especially if one needs to use a two-color interferometer as discussed in the previous section. Secondly, the 400 and 266~nm driving fields improve the energy spacing between adjacent harmonics, making it easier to spatially separate them, for example, by using a monochromator~\cite{Nugent-Glandorf2000,Frassetto2011}. However, the main drawback of shorter wavelength driving fields is that the harmonic cutoff shifts to lower energies due to the reduction of the ponderomotive energy of the electron~\cite{Chang2011,Lin2018}.

\begin{figure}[ht]
 \centering
        \includegraphics[width=0.60\textwidth]{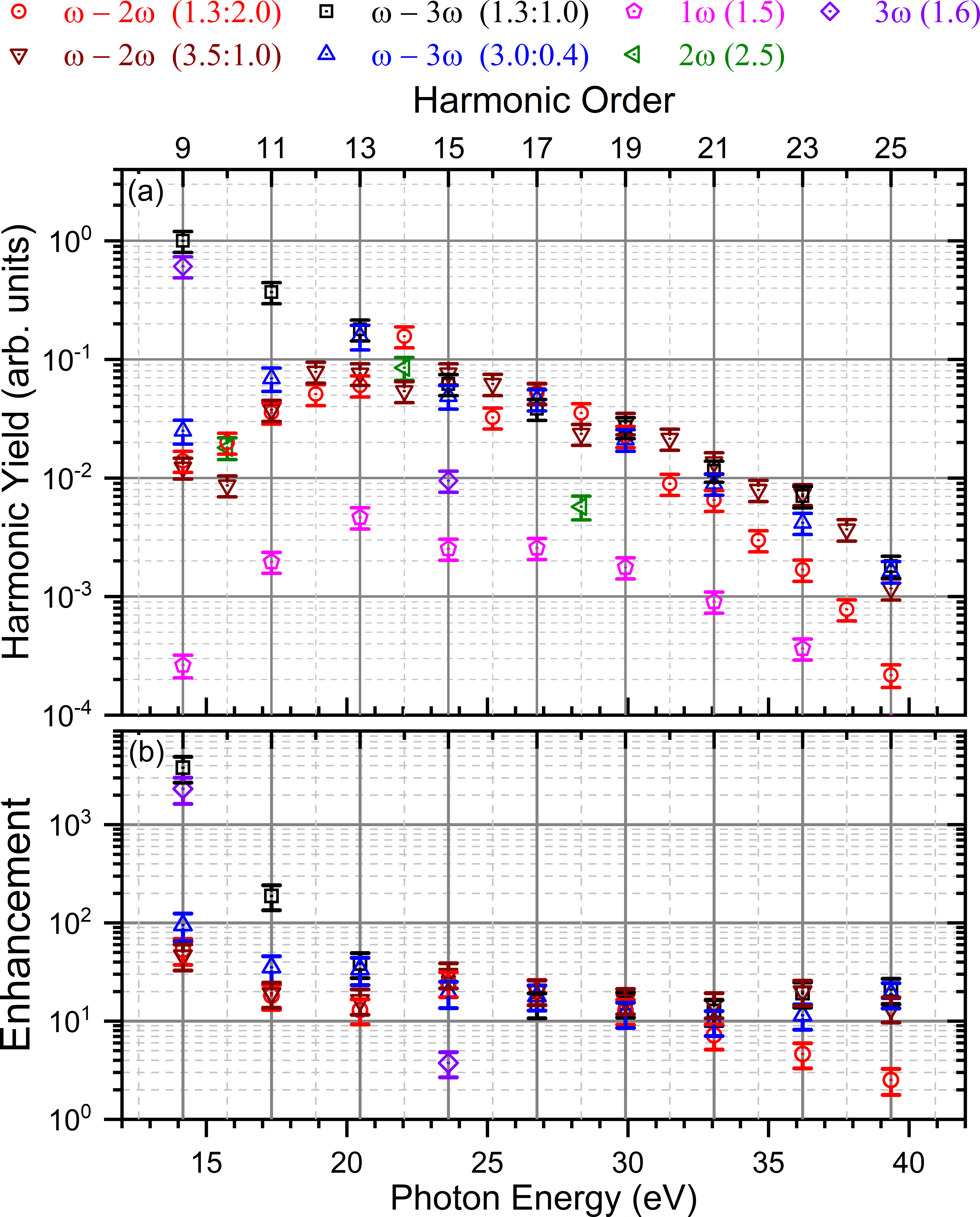}
    \caption{(a) The integrated harmonic yields for the single color 800~nm ($1\omega$), 400~nm ($2\omega$), and 266~nm ($3\omega$) fields as well as the two-color 800-400 ($\omega-2\omega$) and 800-266~nm ($\omega-3\omega$) fields. In the figure legend, the number in parenthesis represents the intensity, in units of $10^{14}$ W/cm$^2$. For the two-color fields, the first number represents the intensity of the fundamental field while the second number is the intensity of the second color. (b) The relative enhancement of the HHG yields over the single-color 800~nm field for similar photon energies. }
    \label{fig:results}
\end{figure}

In Fig.~\ref{fig:results}(a), we show the comparison between the optimized single color 800, 400, and 266~nm driving fields after correcting the detection efficiencies of each individual harmonic. To compare the different single-color fields, we optimize the position of the jet with respect to the focus, thereby changing the phase-matching conditions. In these optimized conditions, the peak intensities of the 800, 400, and 266~nm laser fields are $1.5\times 10^{14}$, $2.5\times 10^{14}$, and $1.6\times 10^{14}$ W/cm$^2$, respectively, with input pulse energies of 2~mJ, 880~$\mu$J, and 310~$\mu$J, respectively. To keep the intensities in the gas medium lower, the gas jet is moved significantly away from the laser's focus, giving the 800 and 400~nm driving fields larger interaction volumes than the 266~nm field.

Even though the 800~nm driving field has more input power, the single-color 400 and 266~nm pulses outperform the 800~nm driving field, as seen in Fig.~\ref{fig:results}(a). Comparing the common photon energies produced by the 800 and 266~nm fields, we find that the photon energies of 14.4 and 24.0~eV are enhanced by factors of approximately 2300 and 4, respectively, when driving HHG with 266~nm as compared to the fundamental 800~nm field, as shown in Fig.~\ref{fig:results}(b). The decrease in yield in HHG driven with 266~nm at 24~eV is expected since this photon energy is beyond the expected cutoff for the 266~nm driving field using the standard estimate of the cutoff photon energy~\cite{Chang2011,Lin2018}.

\section{Two-color driving fields}
In this section, we focus on the results for two-color 800-400~nm and 800-266~nm driving fields. As has been previously shown~\cite{Andiel1999,Cormier2000,Kim2005,Dudovich2006,Liu2006,Kim2008}, mixing $\omega$ and $2\omega$ laser fields breaks the inversion symmetry between half-cycles of the fundamental driving field, leading to the emission of both odd and even harmonics. In contrast, the $\omega-3\omega$ field only produces odd harmonics of the 800~nm driving field since this inversion symmetry between half-cycles of the 800~nm field is preserved.

\begin{figure}[ht!]
 \centering
        \includegraphics[width=0.75\textwidth]{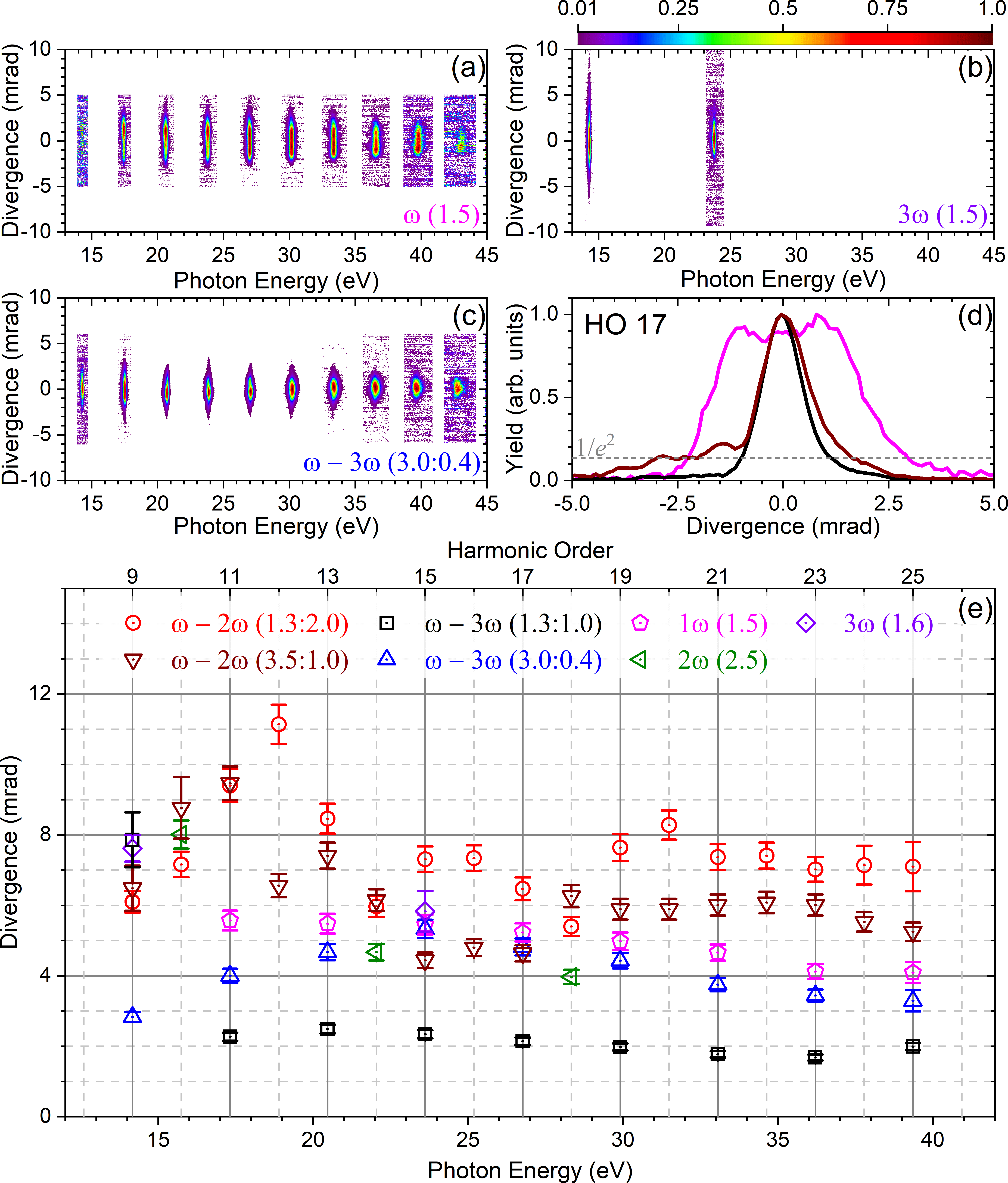}
    \caption{The first three panels show the harmonic yield as a function of divergence angle and photon energy for the optimized (a) 800~nm ($\omega$) field, (b) 266~nm ($3\omega$) field, and the (c) 800-266~nm ($\omega-3\omega$) field with intensity ratio of 3.0:0.4. Note, each individual harmonic is normalized such that the maximum pixel value is 1. (d) The projection of the harmonic 17 onto the divergence axis for the 800~nm, 800-400~nm field with intensity ratio of 3.5:1.0, and 800-266~nm field with intensity ratio of 1.3:1.0. The color coding of the plot follows the legend in panel (e) as well as the Fig.~\ref{fig:results}. Panel (e) shows the divergence for each harmonic for HHG driven by the single- and two-color driving fields. The divergence is defined as the $1/e^2$ full-width measured from the interacting region to the detector plane.}
    \label{fig:diverg}
\end{figure}

In Fig.~\ref{fig:results}(a), we compare the harmonic yields of all single-color and two-color measurements. Furthermore, in Fig.~\ref{fig:results}(b) we compare the relative enhancement of HHG produced by the two-color and 266~nm driving fields with respect to the HHG yields produced by the 800~nm field. Note that for the two-color driving fields, we chose the phase between the two colors that optimizes the total yield for each individual harmonic.

First, concentrating on the 800-400~nm two-color fields, the two intensity ratios we studied consisted of fields with 800 and 400~nm intensities of $1.3$ and $2.0\times10^{14}$ W/cm$^2$, i.e. 1.3:2.0 respectively, as well as $3.5$ and $1.0\times10^{14}$ W/cm$^2$, i.e. 3.5:1.0. Although we do not observe a two-order-of-magnitude enhancement as previously reported~\cite{Kim2005,Kim2008}, we do see a significant increase in photon flux compared to the single color 800~nm driving field. For photon energies less than approximately 35~eV, the two intensity ratios produce about the same photon flux, with an enhancement of $45$ at $14.4$~eV to $16$ at 30~eV. A significant difference between the intensity ratios arises at photon energies larger than $35$~eV, where the ratio of 3.5:1.0 dominates, maintaining about an order of magnitude enhancement over HHG driven by the 800~nm field. We believe that this is simply due to the increased intensity of the 800~nm field, leading to larger ponderomotive energy and cutoff.

In Fig.~\ref{fig:results}(a), we also compare the total HHG yield driven by the two-color 800-400~nm fields to the single-color 400 and 266~nm fields. At similar photon energies below $25$~eV, the single color 400~nm field produces approximately the same flux, within a factor of 2, as the two-color 800-400~nm driving fields. However, at energies similar to $25$~eV, the two-color fields dominate. Again, we speculate that this is due to the enhancement of the cutoff photon energy in the two-color driving fields. On the other hand, the 266~nm driving field outperforms the two-color fields at 14.4~eV by about a factor of 50. However, due to the low cutoff of the 266~nm driving field, the 800-400~nm fields outperform the 266~nm field at 23.6~eV by about a factor of 10.

Switching our focus to the 800-266~nm driving fields, we performed measurements using single-color intensity ratios of 1.3:1.0 and 3.0:0.4, where the intensities are in the $10^{14}$ W/cm$^2$ regime. Here, we observe that the 1.3:1.0 intensity ratio is generally equivalent to or outperforms the 3.0:0.4 intensity ratio, especially at lower photon energies. For example, at 14.4~eV, the 1.3:1.0 intensity ratio leads to approximately a factor of 40 larger harmonic yield than the 3.0:0.4 intensity ratio and 3800 times larger than the 800~nm field. For photon energies above 20~eV, both 800-266~nm intensity ratios perform approximately the same. Both intensity ratios also dominate at all photon energies compared to the single-color 800~nm driving field.

Comparing the 800-266~nm to the 800-400~nm driving fields, we find that for photon energies below $20$~eV, the 1.3:1.0 intensity ratio of the 800-266~nm driving field dominates. For photon energies above 20~eV, both two-color schemes are roughly equivalent, atleast within the estimated errors of our measurements.

Performing a similar comparison between the 800-266~nm and 266~nm driving fields, we find that the two-color field with an intensity ratio of 1.3:1.0 is approximately the same as the single-color 266~nm field at 14.4~eV within our experimental error. On the other hand, the single-color 266~nm driver produced about a factor of 25 more yield than the two-color 800-266~nm driving field with an intensity ratio of 3.0:0.4.

The photon yield is not the only characteristic that many experimentalists may consider when choosing an HHG scheme. In particular, the spatial quality and divergence of the resulting harmonic beam is also important. In Fig.~\ref{fig:diverg}(a-c), we show the harmonic yields as a function of divergence angle and photon energies for the single-color 800, 266, and two-color 800-266~nm field with intensity ratio of 3.0:0.4 in units of 10$^{14}$ W/cm$^2$. Furthermore, in Fig.~\ref{fig:diverg}(e) we show the divergence of each harmonic generated by all the single- and two-color driving schemes we discussed above. Note that we define the divergence as the $1/e^2$ full width, measured from the interacting region to the detector plane. We find that the 800-266~nm driving field generally produces harmonics with smaller divergence angle than the optimized single-color and two-color 800-400~nm driving fields. For some harmonics, such as the 21st harmonic, the divergence for the 800-266~nm field with intensity ratio of 1.3:1.0 is more than two times smaller than the 800~nm field. For example, in Fig.~\ref{fig:diverg}(d), we show the projections of the 17th harmonic for the 800~nm, 800-400~nm field with intensity ratio of 3.5:1.0, and 800-266~nm field with intensity ratio of 1.3:1.0. Note that, to make this plot, we chose the intensity ratio that results in the narrowest distribution for each two-color combination. For the 17th harmonic, we find that the two-color fields produce a much narrower harmonic than the 800~nm driving field. Furthermore, we find that the 800-400~nm field contains some extra structure at larger divergence angles, effectively making the divergence larger than the 800-266~nm case.  In certain applications, having a significantly narrower divergence angle will make the 800-266~nm driving fields preferable to their single- and two-color counterparts.

Our observation that $\omega-3\omega$ fields have narrower divergence angles may suggest that we are selecting which electron trajectories dominate the HHG. Recall that Jin \emph{et al.}~\cite{Jin2014,Jin2014a} theoretically predicted that the $\omega-3\omega$ laser fields produce a large enhancement because these two-color fields optimize the short over the long electron trajectories. Since short electron trajectories usually dominate the total on-axis emitted power~\cite{Brugnera2011}, our observation of significantly narrower divergence angles for the 800-266~nm fields support their prediction.

\section{Summary}
In this paper, we compared the relative flux of HHG driven by the optimized single color 800, 400, and 266~nm driving fields to the two-color 800-400 and 800-266~nm driving fields. We found that, for photon energies below 20~eV, the 800-266~nm driving field with an intensity ratio of 1.3:1.0 significantly outperforms the other driving fields, with the exception of the 266~nm single-color field at 14.4~eV, where both fields produce approximately the same HHG yield. For common photon energies above 20~eV, all two-color driving fields produce approximately the same HHG yield. In addition, we discovered that the 800-266~nm driving field with an intensity ratio of 1.3:1.0 produces harmonics with significantly narrower divergence angle for photon energies above 15~eV compared to all other schemes.

The best generation scheme strongly depends on the application. For example, driving HHG with an 800-400~nm field may not be ideal when using a monochromator~\cite{Nugent-Glandorf2000,Frassetto2011} since the spacing between adjacent harmonics is smaller than the 800-266~nm case. When comparing the two-color driving fields with the single-color 400 and 266~nm fields, we find that these single-color fields are approximately equivalent in photon flux. Therefore, for applications where only these photon energies are needed, it is better to drive HHG with the single-color fields, since they are easier to implement experimentally.

By providing a comparison over a large phase space under the same experimental conditions, we hope that these results can help experimentalists identify which two- or single-color harmonic generation scheme may better fit their experimental needs. It is worth noting that the HHG enhancement that we achieved may be improved upon by further optimizing the phase-matching conditions, for example, using gas cells or waveguides, fine-tuning the intensity ratios, changing the generation medium, or improving the relative phase stability between the two-color fields. We believe that HHG driven by two-color fields is a promising approach which deserves further investigation.

\section*{Funding}
This work and T.S. were partially supported by the National Science Foundation under Award No. IIA-1430493. JRML personnel and operations were supported by the Chemical Sciences, Geosciences, and Biosciences Division, Office of Basic Energy Sciences, Office for Science, U.S. Department of Energy under Award \# DE-FG02-86ER13491. The HITS laser was mainly financed by NSF-MRI grant No: 1229672, with additional contributions from DOD-DURIP grant No. FA2386-12-1-3014 and DOE grant No. DE-FG02-86ER13491. C. T-H  was partially supported by Office of Basic Energy Sciences, Office for Science, U.S. Department of Energy under Award \# DE-SC0024508

\nocite{*}

\bibliography{biblio}

\end{document}